%% file: express-tr.tex
\documentclass{article}

\usepackage{latexsym}
\usepackage{amsmath}
\usepackage{amssymb}
\usepackage{url}
\usepackage{xspace}

\input{include/macros-tr}

\begin{document}

\begin{center}

\textbf{\large Finite-state concurrent programs can be expressed
  pairwise}\\[0.5in]

{\large Paul C. Attie} \\[0.05in]
\vskip 0.05in
{\large Department of Computer Science}\\
{\large American University of Beirut}\\
{\large and}\\
{\large Center for Advanced Mathematical Sciences}\\
{\large American University of Beirut}\\
\texttt{paul.attie@aub.edu.lb}

\today


\end{center}

\begin{abstract}
We present a \intr{pairwise normal form} for finite-state shared memory
concurrent programs: all variables are shared between exactly two processes, and
the guards on transitions are conjunctions of conditions over this pairwise
shared state. This representation has been used to efficiently (in polynomial
time) synthesize and model-check correctness properties of concurrent programs.
Our main result is that any finite state concurrent program can be transformed
into pairwise normal form. Specifically, if $Q$ is an arbitrary finite-state
shared memory concurrent program, then there exists a finite-state
shared memory concurrent program $P$ expressed in pairwise normal form such that
$P$ is strongly bisimilar to $Q$. Our result is constructive: we give an
algorithm for producing $P$, given $Q$.
\end{abstract}



\section{Introduction}
\label{sec:intro}

The \intr{state explosion problem} is recognized as a fundamental
impediment to the widespread application of mechanical finite-state
verification and synthesis methods, in particular, model-checking. The
problem is particularly severe when considering finite-state
concurrent programs, as the individual processes making up such
programs may be quite different (no similarity) and may be only
loosely coupled (leading to a large number of global states).

In previous work \cite{Att99a,AE98,AC04a}, we have suggested a method
of avoiding state-explosion by expressing the synchronization and
communication code for each pair of interacting processes separately
from that for other (even intersecting) pairs. In particular, all
shared variables are shared by exactly one pair of processes.
This ``pairwise normal form'' enables us,
for any arbitrarily large concurrent program,
to model-check correctness
properties for the concurrent compositions of small numbers of
processes (so far 2 or 3) and then conclude that these properties 
also hold in the large program.
If $P$ is a concurrent program consisting of $K$ processes each having
$O(N)$ local states, then we can verify the deadlock freedom of $P$
in $O(K^3 N^3 b)$ time\footnote{$b$ is the maximum branching in the local
  state transition relation of a single process} or $O(K^4 N^4)$ time,
using either of two conservative tests \cite{AC04a},
and we can verify safety and liveness properties of $P$ in $O(K^2 N^2)$ time
\cite{Att99a,AE98}.

A key question regarding the pairwise approach is: does it give up
expressive power? That is, in requiring synchronization and
communication code to be expressed pairwise, do we constrain the set of
concurrent programs that can be represented? In this paper, we answer
this question in the negative: we show that for any concurrent program
$Q$, we can (constructively) produce a concurrent program $P$ that is
in pairwise normal form, and that is strongly bisimilar to $Q$.

The rest of the paper is as follows.
Section~\ref{sec:prelims} presents our model of concurrent computation
and defines the global state transition diagram of a concurrnt
program.  
Section~\ref{sec:pairwise} defines pairwise normal form.
Section~\ref{sec:expressiveness} presents our main result: any finite-state
concurrent program can be expressed in pairwise normal form.
Section~\ref{sec:related} discusses related work, and
Section~\ref{sec:conc} concludes.

%
\section{Technical Preliminaries}
\label{sec:prelims}

\subsection{Model of concurrent computation}
\label{sec:model}

We consider finite-state shared memory concurrent programs of the form 
$P = P_1 \| \cdots \| P_K$ that consist of a finite number $n$ of fixed sequential
processes $P_1, \ldots, P_K$ running in parallel.
Each $P_i$ is a \intr{synchronization skeleton} \cite{EC82}, that is,
a directed multigraph where each node is a (local)
state of $P_i$ (also called an \intr{$i$-state} and is labeled by a unique name ($s_i$), and where each
arc is labeled with a guarded command \cite{Dij76} $B_i \ar A_i$
consisting of a guard $B_i$ and corresponding action $A_i$.
Each node must have at least one outgoing arc, i.e., a skeleton contains no ``dead ends.''
With each $P_i$, we associate a set $\AP_i$ of \intr{atomic
propositions}, and a mapping $V_i$ from local states of $P_i$ to
subsets of $\AP_i$: $V_i(s_i)$ is the set of atomic propositions that
are true in $s_i$. As $P_i$ executes transitions and changes its
local state, the atomic propositions in $\AP_i$ are updated.
Different local states of $P_i$ have different truth assignments:
$V_i(s_i) \ne V_i(t_i)$ for $s_i \ne t_i$.
Atomic propositions are not shared: $\AP_i \ints \AP_j = \emptyset$
when $i \ne j$.
Other processes can read (via guards) but not update the atomic
propositions in
$\AP_i$.
We define the set of all atomic propositions $\AP = \AP_1 \un \cdots \un \AP_K$.
There is also a set $\SH = \{x_1,\ldots,x_m\}$
of shared variables, which can be
read and written by every process.
These are updated by the action $A_i$.
A {\em global state} is a
tuple of the form $(s_1, \ldots, s_K,v_1,\ldots, v_m)$ where
$s_i$ is the current local state of $P_i$ and $v_1,\ldots,v_m$ is a list
giving the current values of $x_1,\ldots,x_m$, respectively.
A guard $B_i$ is a predicate on global states, and so can
reference any atomic proposition and any shared variable.
An action $A_i$ is any
piece of terminating pseudocode 
that updates the shared variables.\footnote{We will only use
straight-line code in this paper, so termination is always guaranteed.}
We write just $A_i$ for $\ltrue \ar A_i$ and just $B_i$ for $B_i \ar skip$,
where $skip$ is the empty assignment.


We model parallelism as usual by the nondeterministic
interleaving of the ``atomic" transitions of the individual
processes $P_i$.  Let
$s = (s_1, \ldots , s_i, \ldots, s_K, v_1,\ldots, v_m)$
be the current global state, and let $P_i$ contain an arc from node $s_i$ to
$s'_i$ labeled with $B_i \ar A_i$. We write such an arc as the tuple
$(s_i, B_i \ar A_i, s'_i)$, and call it a $P_i$-\intr{arc} from $s_i$ to $s'_i$.
We use just \intr{arc} when $P_i$ is specified by the context.
If $B_i$ holds in $s$, then a permissible next state is
$s' = ( s_1, \ldots, s'_i, \ldots, s_K, v'_1,\ldots, v'_m )$ where
 $v'_1, \ldots, v'_m$ are the
new values for the shared variables resulting from action
$A_i$.
Thus, at each step of the computation, a process with an enabled arc is
nondeterministically selected to be executed next.
The \intr{transition relation} $R$ is the set of all such 
$(s,i,s')$.
The arc from node $s_i$ to $s'_i$ is {\em enabled} in state $s$.
An arc that is not enabled is {\em blocked}.
Our model of computation is a high-atomicity model, since a process $P_i$
can evaluate the guard $B_i$, execute the action $A_i$, and change its
local state, all in one action.

Recall that we define a global state to be a tuple of local states and
shared variable values, rather than a ``name'' together with a
labeling function $L$ that gives the associated valuation, 
A consequence of this definition is that two different global states
must differ in either some local state or some shared variable value.
Since we require different local states to differ in at least one
atomic proposition value, we conclude that two different global states
differ in at least one atomic proposition value or one shared variable
value.

We define the valuation corresponding to a global state 
$s = (s_1, \ldots , s_i, \ldots, s_K,$\\$v_1,\ldots, v_m)$ as 
follows.
For an atomic proposition $p_i \in \AP_i$:
$s(p_i) = \ltrue$ if $p_i \in V_i(s_i)$, and
$s(p_i) = \lfalse$ if $p_i \not\in V_i(s_i)$.
For a shared variable $x_\l$, $1 \le \l \le m$: $s(x_\l) = v_\l$.
We define $s \pj \AP$ to be the set $\{ p \in \AP ~|~ s(p) = \ltrue\}$
i.e., the set of propositions that are true in state $s$.
$s \pj \AP$ is essentially the projection of $s$ onto the atomic propositions.
Also, $s \pj i$ is defined to be $s_i$, i.e., the local state of $P_i$ in $s$.
We also define $s \pj \SH$ to be the set $\{ \tpl{p, s(x)} ~|~ x \in \SH
\}$, i.e., the set of all pairs consisting of a shared variable $x$ in $\SH$
together with the value that $s$ assigns to $x$.

Let $St$ be a given set of initial states in which computations of
$P$ can start.
A \emph{computation path} is a sequence of states
whose first state is in $St$ and 
where each successive pair of states is related by $R$.
A state is \intr{reachable} iff it lies on some computation path.
%
Since we must specify the start states $St$ in order for the 
computation paths to be well-defined, we re-define our notion of a program 
to be $P = (St, P_1 \pl \cdots \pl P_K)$, i.e., a program consists of
the parallel composition of $K$ processes, together with a set $St$ of 
initial states.


For technical convenience, and without loss of generality, 
we assume that no synchronization skeleton
contains a node with a self-loop. The functionality of a self-loop
(e.g., a busy wait) can always be achieved by using a loop containing
two local states.
Thus, a transition by $P_i$ changes the local state of $P_i$, and
therefore the value of at least one atomic proposition in $\AP_i$. 
Hence, no global state $s$ has a self loop, i.e., a transition by some
$P_i$ both starting and finishing in $s$.

For a local state $s_i$, define $\stof{s_i}$ as follows:
\begin{definition}[State-to-Formula Translation]
\label{def:stof}
\[
\stof{s_i}  =  ``(\AND_{p \in V_i(s_i)} p) \;\ \land\;\
                 (\AND_{p \not\in V_i(s_i)} \neg p)\mbox{\rm ''}
\]
         \mbox{where $p$ ranges over $\AP_i$.}
\end{definition}
$\stof{s_i}$ converts a local state $s_i$ into a propositional formula over $\AP_i$.

If $s$ is a global state and $B$ is a guard, we define
$s(B)$ by the usual inductive scheme:
$s($``$x = c$"$) = \ltrue$ iff $s(x) = c$,
$s(B1 \land B2) = \ltrue$ iff $s(B1) = \ltrue$ and $s(B2) = \ltrue$,
$s(\neg B1) = \ltrue$ iff $s(B1) = \lfalse$.
If $s(B) = \ltrue$, we also write $s \sat B$.

\subsection{The Global State Transition Diagram of a Concurrent Program}
\label{subsec:gstd}

\bd [Global state transition diagram]
\label{def:gstd}
Given a concurrent program $P = P_1 \| \cdots \| P_K$ and a set $St$
of initial global states for $P$, the \intrdef{global state
transition diagram generated by $P$} is a Kripke structure 
$M = (St, S, R)$ given as follows:
(1) $S$ is the smallest set of global states satisfying 
   (1.1) $St \sub S$ and 
   (1.2) if there exist $s \in S, i \in \onetok$\footnote{We use
   $\onetok$ for the set consisting of the natural numbers $1,\ldots,K$.}, and $u$ 
    such that 
            $(s,i,u)$ is in the next-state relation defined above in
            Section~\ref{sec:model}, then
         $u \in S$,
and 
(2) $R$ is the next-state relation restricted to $S$.
\ed

We define strong bisimulation in the standard way.

\bd[Strong Bisimulation]
\label{def:bisim}
Let $M = (St, S, R)$ and $M' = (St', S', R')$ be two Kripke
structures with the same underlying set $\AP$ of atomic propositions.
A relation $B \sub S \times S'$ is a \intrdef{strong  bisimulation}
iff:
\bn

\item \label{def:bisim:ap}
if $B(s,s')$ then $s \pj \AP = s' \pj \AP$

\item \label{def:bisim:left-trans}
if $B(s,s')$ and $(s,i,u) \in R$ then
           $\ex u': (s',i,u') \in R' \land B(u,u')$

\item \label{def:bisim:right-trans}
if $B(s,s')$ and $(s',i,u') \in R$ then
           $\ex u: (s,i,u) \in R \land B(u,u')$

\en

We also define $\sim$ to be the union of all strong bisimulation
relations:\\
 $\sim \;=\; \UN \{B : B \mbox{ is a strong bisimulation}\}$.

We say that $M$ and $M'$ are \intrdef{strongly bisimilar},
and write $M \sim M'$,  if and only if
there exists a strong bisimulation $B$ such that 
$\fa s \in St, \ex s' \in St': B(s's')$ and
$\fa s' \in St', \ex s \in St: B(s's')$.
\ed

\section{Pairwise normal form}
\label{sec:pairwise}

Let $\gd, \gc$ be binary infix operators.
A \intr{general guarded command} \cite{AE98} is either a guarded command as given in
Section~\ref{sec:model} above, or has the form
$G_1 \gd G_2$ or $G_1 \gc G_2$, where $G_1$, $G_2$ are general guarded commands.
Roughly, the operational semantics of
$G_1 \gd G_2$ is that either $G_1$ or $G_2$, but not both, can be executed,
and the operational semantics of
$G_1 \gc G_2$ is that both $G_1$ or $G_2$ must be executed, that is, the guards
of both $G_1$ and $G_2$ must hold at the same time, and the bodies of
$G_1$ and $G_2$ must be executed simultaneously, as a single parallel assignment
statement. For the semantics of $G_1 \gc G_2$ to be well-defined, there must be
no conflicting assignments to shared variables in $G_1$ and $G_2$. This will
always be the case for the programs we consider. We refer the reader to
\cite{AE98} for a comprehensive presentation of general guarded commands.

\bd[Pairwise Normal Form]
\label{def:pairwise}
A concurrent program $P = P_1 \| \cdots \| P_K$ is in
\intr{pairwise normal form} iff the following four conditions all hold:
\bn

\item \label{def:pairwise:arc}
 every arc $a_i$  of every process $P_i$ has the form\\
$a_i =  (s_i,
         \gc_{j \in I(i)} \gd_{\l \in \{1,\ldots,n_j\}} \CB{i}{j}{\l} \ar \CA{i}{j}{\l},
         t_i)$,
where 
    $\CB{i}{j}{\l} \ar \CA{i}{j}{\l}$ is a guarded command,
    $I$ is an irreflexive symmetric relation over $\onetok$ that defines a
    ``interconnection'' (or ``neighbors'') relation amongst processes, and 
    $I(i) = \{j ~|~ (i,j) \in I\}$,

\item \label{def:pairwise:shvar}
 variables are shared in a pairwise manner, i.e., for each $(i,j) \in I$, there is some set
    $\SH_{ij}$ of shared variables that are the only variables that
    can be read and written by both $P_i$ and $P_j$, 

\item \label{def:pairwise:guard}
$\CB{i}{j}{\l}$ can reference only variables in $\SH_{ij}$
    and atomic propositions in $\AP_j$,  and 

\item \label{def:pairwise:assig}
$\CA{i}{j}{\l}$ can update only variables in $\SH_{ij}$.
\en
\ed

For each neighbor $P_j$ of $P_i$,
$\gd_{\l \in [1:n]} \CB{i}{j}{\l} \ar \CA{i}{j}{\l}$ specifies
$n$ alternatives $\CB{i}{j}{\l} \ar \CA{i}{j}{\l}$, $1 \le \l \le n$ for the
interaction between $P_i$ and $P_j$ as $P_i$ transitions from $s_i$ to $t_i$.
$P_i$ must execute such an interaction with each of its neighbors in order to
transition from $s_i$ to $t_i$.
We emphasize that $I$ is not necessarily the set of all pairs, i.e.,
there can be processes that do not directly interact by reading each
others atomic propositions or reading/writing pairwise shared variables.
We do not assume, unless otherwise stated, that processes are isomorphic,
or ``similar.''

We use a superscript $I$ to indicate the relation $I$,
e.g., process $P_i^I$, and  $P_I^i$-arc $\mv{i}{I}$.
We define $\mv{i}{I}.start = s_i$,
$\mv{i}{I}.guard_j = \bigor_{\l \in \{1,\ldots,n_j\}} \CB{i}{j}{\l}$, and
$\mv{i}{I}.guard = \ba_{j \in I(i)} a_i.guard_j$.
If $P^I = P_{1}^I \pl \ldots \pl P_{K}^I$ is a concurrent program with interconnection relation $I$, then
we call $P^I$ an \intr{$I$-system}.
For the special case when 
$I = \{ (i,j) ~|~ i,j \in \onetok, i \ne j\}$,
i.e., $I$ is the complete interconnection relation, we omit the
superscript $I$.

In pairwise normal form, the synchronization code for $\PP{i}{I}$ with
one of its neighbors  $\PP{j}{I}$ (i.e.,
$\gd_{\l \in \{1,\ldots,n_j\}} \CB{i}{j}{\l} \ar \CA{i}{j}{\l}$) 
is expressed separately from the synchronization code for $\PP{i}{I}$
with
another neighbor $P_k^I$ (i.e., 
$\gd_{\l \in \{1,\ldots,n_k\}} \CB{i}{k}{\l} \ar \CA{i}{k}{\l}$)
We can exploit this property  to define ``subsystems'' of an $I$-system
$P$ as follows.
Let $J \sub I$ and $range(J) = \{i ~|~ \ex j: (i,j) \in J\}$.
If $\mv{i}{I}$ is a arc of $\PP{i}{I}$ then define
$\mv{i}{J} = (s_i,
         \gc_{j \in J(i)} \gd_{\l \in \oneton} \CB{i}{j}{\l} \ar \CA{i}{j}{\l},
         t_i)$.
Then the \intr{$J$-system} $P^J$ is 
$P_{j_1}^J \pl \ldots \pl P_{j_n}^J$ where
$\{j_1,\ldots,j_n\} = range(J)$ and 
$\PP{j}{J}$ consists of the arcs
$\{ \mv{i}{J} ~|~ \mbox{ $\mv{I}{I}$ is a arc of $\PP{j}{I}$}\}$.
Intuitively, a $J$-system consists of the processes in $range(J)$,
where each process contains only the synchronization code needed for
its $J$-neighbors, rather than its $I$-neighbors.
If $J = \{ \{i,j\}\}$ for some $i,j$ then $P_J$ is a \intr{pair-system},
and if
$J = \{ \{i,j\}, \{j,k\} \}$ for some $i,j,k$ then $P_J$ is a \intr{triple-system}.
For $J \sub I$, 
$M_J = (St_J,S_J,R_J)$ is the GSTD of $P^J$ as defined in
Section~\ref{sec:model}, and
a global state of $P^J$ is a \intr{$J$-state}.
If $J = \{ \{i,j\}\}$, then we write 
$M_{ij} = (St_{ij},S_{ij},R_{ij})$ instead of $M_J = (St_J,S_J,R_J)$.


In \cite{Att99a,AE98,Att03a} we give, in pairwise normal form,
solutions to many well-known problems, such as dining philosophers,
drinking philosophers, mutual exclusion, $k$-out-of-$n$ mutual
exclusion, two-phase commit, and replicated data servers.  We
conjecture that any finite-state concurrent program can be rewritten
(up to strong bisimilation) in pairwise normal form.  The restriction
to pairwise normal form enables us to mechanically verify certain
correctness properties very efficiently.  Recall that $K$ is the
number of processes, $b$ is the maximum branching in the local state
transition relation of a single process, and $N$ is the size of the
largest process.  Then, safety and liveness properties that can be
expressed over pairs of processes can be verified in time $O(K^2 N^2)$
by model-checking pair-systems, \cite{Att99a,AE98}, and
deadlock-freedom can be verified in time in $O(K^3 N^3 b)$ or $O(K^4
N^4)$ using either of two conservative tests \cite{AC04a}, which in
turn operate by model checking triple-systems.  Exhaustive state-space
enumeration would of course require $O(N^K)$ time.

\section{The Pairwise Expressiveness Result}
\label{sec:expressiveness}





Let $Q = (St_Q, Q_1 \| \cdots \| Q_K)$ be an arbitrary finite-state shared memory
concurrent program as defined in Section~\ref{sec:model} above,
with each process $Q_i$ having an associated set $\AP_i$ of atomic
propositions and with shared variables $x_1,\ldots,x_m$. 
The transformation of $Q$ to pairwise normal form proceeds in three
phases, as given in the sequel.

\subsection{Phase One}
\label{sec:phase-one}

First, we generate $M_Q$, the GSTD of $Q$, as given by Definition~\ref{def:gstd}.
By construction of Definition~\ref{def:gstd}, all states in $M_Q$ are reachable.
We then execute the algorithm given in Figure~\ref{fig:unique-in} on
$M_Q$ which transforms $M_Q$ intro a Kripke structure 
$M'_Q = (St'_Q, S'_Q, R'_Q)$
which is bisimilar to $M_Q$ and which has the property that 
all incoming transitions into a state are labeled with the same process index.
%
%
This is not strictly necessary, but significantly simplifies the
transformation to pairwise normal form.

Define $\inprocs(s) = \{i \in \onetok ~|~ \ex s' : (s',i,s) \in R_Q \}$.
We also introduce a new shared variable $in$ whose value in a state $s$
will be the process index that labels the transitions incoming into $s$.

\bfg
\horline

\vspace{-2ex}

\begin{tabbing}
   aa\=aaa\=aa\=aaa\=aaa\=aa\= \kill
\>$\trans(M_Q,M'_Q)$\\[1.0ex]
\>$St'_Q := St_Q$; $S'_Q := S_Q$; $R'_Q := R_Q$;\\
\>\REPEAT until there is no change in $M'_Q$\\
\>   \>let $s$ be a state in $M'_Q$ such that $|\inprocs(s)| > 1$;\\
\>   \>\FORALL $i \in \inprocs(s)$ \DO\\
\>   \>   \>create a new marked state $s^i$ such that 
				$s^i \up \AP = s \up \AP$,
				$s^i \up \SH = s \up \SH$\\
\>   \>   \>\IF $s \in St_Q$ \THEN $\setinsert{St'_Q}{s^i}$ \ENDIF;\\
\>   \>   \>$\setinsert{S'_Q}{s^i}$;\\
\>   \>   \>\FORALL $j,u : (s,j,u) \in R_Q$ \DO 
                              $\setinsert{R'_Q}{(s^i,j,u)}$ \ENDFOR;\\
\>   \>   \>\FORALL $u: (u,i,s) \in R_Q$ \DO 
                              $\setinsert{R'_Q}{(u,i,s^i)}$ \ENDFOR;\\
\>   \>   \>$\setdelete{St'_Q}{s}$;\\
\>   \>   \>$\setdelete{S'_Q}{s}$;\\
\>   \>   \>remove all transitions incident on $s$ from $R'_Q$\\
\>   \>\ENDFOR\\
\>\ENDREPEAT
\end{tabbing}

\vspace{-3ex}

\horline
\caption{Transformation of $M_Q$ so that all incoming
         transitions are labeled with the same process index.}
\label{fig:unique-in}
\efg

\bp
\label{prop:unique-term}
Procedure $\trans$ terminates.
\ep
\bpr
Each iteration of the \REPEAT loop (line 2) reduces the number of
states $s$ such that $|\inprocs(s)| > 1$ by one. Since $M'_Q$ is
initially set to $M_Q$, which is finite, this cannot go on forever.
\epr

\bp
\label{prop:unique-bisim}
$M'_Q \sim M_Q$ is a loop invariant of the \REPEAT loop (line 2) of
$\trans$.  
\ep

\textit{Proof of Proposition~\ref{prop:unique-bisim}}.
\bpr
Let $n_0$ be the number of iterations that the \REPEAT loop executes.
Let $M^n = (St^n, S^n, R^n)$ be the value of $M'_Q$ at the end
of the $n$'th iteration, (for all $n \le n_0$) with $M^0$ being the
initial value $M_Q$. We will also use the superscript $n$ for states
in $M^n$, when needed.
We show that $\fa n: 0 < n \le n_0: M^{n-1} \sim M^{n}$.

Consider the $n$'th iteration of the $\REPEAT$ loop.  In this
iteration, $M^n$ results from $M^{n-1}$ by deleting some 
state $s$ and adding some states $s^{i_1} \ldots s^{i_\l}$,
where $\set{i_1,\ldots i_\l} = \inprocs(s)$.
Since each of $s^{i_1} \ldots s^{i_\l}$ have the same successor states
as $s$, and agree with $s$ on the values of all atomic propositions,
we have $s \sim s^{i_1}, \ldots, s \sim s^{i_\l}$.
Let $u$ be an arbitrary predecessor of $s$ in $M^{n-1}$,
i.e., $(u^{n-1},j,s) \in R^{n-1}$, where $u^{n-1}$ indicates the
occurrence of $u$ in $M^{n-1}$.
At the end of the iteration, we have $(u^n,j,s^j) \in R^n$.
Since $s \sim s^j$, we have $u^{n-1} \sim u^n$, i.e.,
the occurrence of $u$ in $M^{n-1}$ is bisimilar
to the occurrence of $u$ in $M^n$.
Since all other states in $M^{n-1}$ and $M^n$ have an unchanged set of
successors, we conclude that $M^{n-1} \sim  M^n$.

By a straightforward induction on $n$, and using the transitivity of $\sim$,
we can show that
$\fa n: 0 < n \le n_0: M^0 \sim M^{n}$.
Thus $M^0 = M^{n_0}$.
Now $M_Q = M^0$ and $M'_Q = M^{n_0}$, and the proposition is established.
\epr

\bp
\label{prop:unique-correct}
Upon termination of procedure $\trans$,\\
(1) $M'_Q \sim M_Q$, and\\
(2) every state $s$ in $M'_Q$ satisfies $|\inprocs(s)| \le 1$.
\ep
\bpr
(1) follows from Proposition~\ref{prop:unique-bisim}.
(2) follows immediately fom inspecting line 2 of procedure $\trans$.
\epr

For all $s \in S'_Q$ such that $|\inprocs(s)| = 1$,
define $\inproc(s)$ to be the unique $i$ such that 
$\ex s' : (s',i,s) \in R'_Q$.

\bp
\label{prop:unique-global}
Upon termination of procedure $\trans$, for any two states $s,u$ in
$M'_Q$,
$s \pj \AP \ne u \pj \AP$ or
$s \pj \SH \ne u \pj \SH$ or
$in(s) \ne in (u)$.
\ep
\bpr
Immediate by construction of procedure $\trans$.
\epr

\subsection{Phase Two}
\label{sec:phase-two}

We exploit the unique incoming process index property of $M'_Q$ to
extract a program $P = (St_P, P_1 \| \cdots \| P_K)$ from $M'_Q$
such that $P$ is bisimilar to 
$Q = (St_Q, Q_1 \| \cdots \| Q_K)$ and $P$ is in pairwise normal form.
The interconnection relation $I$ for $P$ is the complete relation, and
so we omit the superscripts $I$ on $P$ and $P_i$. 
$P$ operates by emulating the execution of $Q$. 
In the sequel, let $i,j,k$ implicitly range over $\onetok$, with
possible further restriction, e.g., $i \ne j$.
With each process $P_i$ we associate the following state variables, with the indicated
access permissions and purpose
\be

\item \empb{The atomic propositions in $\AP_i$}. These are written by $P_i$ and read by
   all processes.  For each process $P_i$, these enable $P_i$ to
   emulate the local state of $Q_i$, which is defined by the same set
   $\AP_i$ of atomic propositions.

\vspace{1.5ex}

\item \empb{A shared variable $x_{ij}^i$ for every $x \in \SH$ and $j \in
   \onetok$}. These are written by $P_i$ and read by $P_j$. These enable $P_i$ to
   emulate the updates that $Q_i$ makes to $x$.  When $P_i$ is the
   last process to have executed, any other process $P_j$ will read
   $x_{ij}^i$ to find the correct emulated value of $x$, since this
   value will have been computed by $P_i$ and stored in $x_{ij}^i$ for
   all $j \in \onetok$.
   For technical convenience, we admit $x_{ii}^i$.
   We select some $\l \in \onetok - \{i\}$ arbitrarily and define 
   $x_{ii}^i$ to be shared pairwise between $P_i$ and $P_\l$. 
   This is needed to conform technically to
   Definition~\ref{def:pairwise}.
   $P_\l$ will not actually reference $x_{ii}^i$.

\vspace{1.5ex}

\item \empb{A timestamp $t_i^j$ for every $j \in \onetok$}.
   These are written and read by $P_i$ only.  Timestamps have values in
   $\set{0,1,2}$. We define orderings $\tls$, $\tgt$ on timestamps as
   follows \cite{DS97}: $0 \tls 1$, $1 \tls 2$, and $2 \tls 0$, and $t
   \tgt t'$ iff $t' \tls t$.  Note that $\tls$ is not transitive.  The
   purpose of $t_i^j$ and $t_j^i$ is to enable the pair of processes
   $P_i$ and $P_j$ to establish an ordering between themselves by
   computing $t_i^j \tls t_j^i$.  If $t_i^j \tgt t_j^i$, then $P_i$
   executed a transition more recently than $P_j$, and vice-versa.
   The timestamp $t_i^i$ is
   unused, so we do not worry about initializing it, or what is value
   is in general.

\vspace{1.5ex}

\item A \empb{timestamp vector $tv_{ij}^i$ for every $j \in \onetok$.}
   A $K$-tuple whose value is maintained equal to
   $\tpl{t_i^1,\ldots,t_i^K}$.  It is written by $P_i$ and read by
   $P_i$ and $P_j$.  Its purpose is to allow $P_i$ to communicate to
   $P_j$ the values of $P_i$'s timestamps w.r.t. all other
   processes. By reading all  $tv_{ij}^i$, $i \in \onetok - \{j\}$,
   process $P_j$ can correctly infer the index of the last process to
   execute. This allows $P_j$ to read the correct emulated values of
   all shared variables.
   We use $tv_{ij}^i.k$ to denote the $k$'th element of $tv_{ij}^i$, 
   which is the value of $t_i^k$. 
   For technical convenience, we admit $tv_{ii}^i$.
   We select some $\l \in \onetok - \{i\}$ arbitrarily and define 
   $tv_{ii}^i$ to be shared pairwise between $P_i$ and $P_\l$. 
   This is needed to conform technically to
   Definition~\ref{def:pairwise}.
   $P_\l$ will not actually reference $tv_{ii}^i$.

\ee
For all the above, the order of subscripts does not matter, e.g., 
$tv_{ij}^i$ and $tv_{ji}^i$ are the same variable, etc.

The essence of the emulation is to deal correctly with the shared
variables. This depends upon every process being able to compute the
index of the last process to execute, as described above.
Define the auxiliary (``ghost'') variable $\last$ to be the index of
the last process to make a transition.
As described above, every process $P_j$ can compute the value of
$\last$ ($\last$ is not explicitly implemented, since doing so would
violate pairwise normal form).  Then, $P_j$ reads the variable
$x_{\last,j}^\last$ that it shares with $P_{\last}$ to find an up to
date value for the variable $x$ in $Q$.  Together with the unique
incoming process index property of $M'_Q$, this allows $P_j$ to
accurately determine the currently simulated global state of $M'_Q$.
$P_j$ can then update its associated shared variables and atomic
propositions to accurately emulate a transition in $M'_Q$.

Let $M_P$ be the GSTD of $P$, as given by Definition~\ref{def:gstd}.
We will define $P = (St_P, P_1 \| \cdots \| P_K)$ so that 
$M'_Q$ and $M_P$ are bisimilar.

We start with $St_P$.
For each initial state $u_0$ of $M'_Q$, we create a corresponding
initial state $r_0 \in St_P$ so that:
\bleqn{}
$r_0 \pj \AP = u_0 \pj \AP$
\eleqn
\vspace{-0.4in}
\bleqn{}
$\AND_{x \in \SH, i,j} r_0(x_{ij}^i) = u_0(x)$
\eleqn
%
%
Now for the bisimulation between $M'_Q$ and $M_P$ to work properly,
we will require that $\inproc(u) = s(\last)$, where $u,s$ are
bisimilar states of $M'_Q$, $M_P$, respectively.
It is possible, however, that some initial state $u_0$ of $M'_Q$ does not
have an incoming transition, and so $\inproc(u_0)$ is undefined.
We deal with this as follows.

Call an initial state (of either $M'_Q$ or $M_P$) that does not have an
incoming transition a \intr{source state}.
Since we defined the corresponding $r_0$ above so that
$x_{ij}^i$ has the correct value (namely $u_0(x)$) for all
$i,j$, we can let any process be the ``last'', as determined by the
timestamps. Thus, for a source state $u_0$ in $M'_Q$ and its
corresponding source state $r_0$ in $M_P$, we set:
\[
r_0(t_i^j) = \left\{ 
\begin{array}{l}
1  ~\mathrm{if}~ i  =  1 \land j \ne 1 \\
0  ~\mathrm{if}~ i \ne 1 \land j  =  1 \\
X ~\mathrm{if}~  i \ne 1 \land j \ne 1\\
\end{array} 
\right.
\]
where $X$ denotes a ``don't care,'' i.e., any value in $\set{0,1,2}$
can be used.
This has the effect of making $P_1$ the ``last'' process to have executed in
a source state, i.e., setting $r_0(\last) = 1$.
We now extend the definition of $\inproc$ to source states by defining
$\inproc(u_0) = 1$ for every source state $u_0 \in St'_Q$. 
Together with the fact that states in $M'_Q$ are uniquely determined
by the atomic proposition and shared variable values, this
automatically takes care of the bisimulation matching between source
states in $M'_Q$ and source states in $M_P$, without the need for an
extra case analysis.
Note also that $\inproc(u)$ is now defined for all states $u$ in $M'_Q$.

For an initial state $u_0$ of $M'_Q$ that is not a source state, 
and its corresponding initial state $r_0$ in $M_P$, we set:
\[
r_0(t_i^j) = \left\{ 
\begin{array}{l}
1 ~\mathrm{if}~ i  =  \inproc(u_0)  \land  j \ne \inproc(u_0) \\
0 ~\mathrm{if}~ i \ne \inproc(u_0)  \land  j  =  \inproc(u_0) \\
X ~\mathrm{if}~ i \ne \inproc(u_0)  \land  j \ne \inproc(u_0) \\
\end{array} 
\right.
\]
where again $X$ means ``don't care.'' This has the effect of setting
$r_0(\last) = \inproc(u_0)$, as required.

For all initial states  $r_0 \in St_P$, whether thay are source states
or not, we set the timestamp vector values so that:
\bleqn{}
$\AND_{i,j,k} r_0(tv_{ij}^i.k) = r_0(t_i^k)$
\eleqn


For each transition $(u,i,v)$ in $M'_Q$, we generate a single arc
$ARC_i^{u,v}$ in $P_i$ as follows.
$ARC_i^{u,v}$ 
starts in local state $u \pj i$ of $P_i$ and ends in local
state $v \pj i$ of $P_i$.
Let $\inproc(u) = c$. Then
the guard $B_i^{u,v}$ of $ARC_i^{u,v}$ is defined as follows:
\bleqn{}
     $B_i^{u,v} \;\df\;
       (\last = c) \;\land\;
       \AND_{j \ne i} \stof{u \pj j} \;\land\;
       (\AND_{x \in \SH} x_{ci}^c = u(x))$
\eleqn
The first conjunct checks that the last process that executed is the
process with index $\inproc(u)$.
The second conjunct checks that all atomic propositions have the
values assigned to them by global state $u$.
The third conjunct checks that all shared variables have the
values assigned to them by global state $u$.


The action $A_i^{u,v}$ of $ARC_i^{u,v}$ is defined to be 
\begin{eqnarray*}
\pl_{j \ne i} \ t_i^j := \step(t_i^j, tv_{ji}^i.j); \\       
\pl_{j} \ tv_{ij}^i := \tpl{t_i^1,\ldots,t_i^K};\\
\pl_{j, x \in \SH} \ x_{ij}^i := v(x)         
\end{eqnarray*}
where $\step(t,t')$ is given in Figure~\ref{fig:step}.
This cannot be factored into pairwise actions  $\CA{i}{j}{m}$ because
all the $t_i^j$ are used to update all the $tv_{ij}^i$. The solution
is to make the $t_i^j$ part of the local state of $P_i$. We do this in
phase 3 below. For now, we show that program $P$ with the arcs given
by $ARC_i^{u,v} = (u \pj i, B_i^{u,v} \ar A_i^{u,v}, v \pj i)$
is bisimilar to program $Q$.


\bfg
\horline
\begin{tabbing}
\=aaa\=aaa\=aaa\=aaa\=aa\= \kill
\>$\step(t, t')$\\[0.5ex]
\>Precondition: $0 \le t, t' \le 2$, that is, $t,t'$ are timestamp values\\
\>\IF $t \tgt t'$ \THEN $\RETURN(t)$\\
\>\ELSE\\
\>   \>  \>\IF $t = 0 \land t' = 1$ \THEN $\RETURN(2)$ \ENDIF;\\
\>   \>  \>\IF $t = 1 \land t' = 2$ \THEN $\RETURN(0)$ \ENDIF;\\
\>   \>  \>\IF $t = 2 \land t' = 0$ \THEN $\RETURN(1)$ \ENDIF;\\
\>\ENDIF\\
\end{tabbing}
\horline
\caption{The $\step$ procedure.}
\label{fig:step}
\efg

\bp
\label{prop:invariants}
The following are invariants of $P$:
\bn
\item $\AND_{i, j, k \ne i} tv_{ij}^i.k = t_i^k$
\item $\AND_i ( (\last = i) \;\equiv\; \AND_{j \ne i} t_i^j \tgt  t_j^i )$
\item $\AND_{i,j,k} x_{ij}^i = x_{ik}^i$
\en
\ep
\bpr
By construction of $P$: $St_P$ is defined so that the initial states
all satisfy the above, and the actions  $A_i^{u,v}$ of every process $P_i$ of $P$ are
defined so that their execution preserves the above.
\epr

\bd
\label{def:correctness-sim}
Define $\bsim \sub S'_Q \times S_P$ as follows.
For $u \in S'_Q$, $r \in S_P$, $u \bsim r$ iff:
\bn

\item \label{def:correctness-sim:ap}
$u \pj \AP = r \pj \AP$

\item \label{def:correctness-sim:last}
$\inproc(u) = r(\last)$

\item \label{def:correctness-sim:shvar}
$\AND_{x \in \SH, k} r(\last) = k \imp (\AND_i u(x) = r(x_{ki}^{k}))$

\en
\ed

\vspace{1ex}

\bt
\label{thm:bisim}
$\bsim$ is a strong bisimulation
\et

\textit{Proof of Theorem~\ref{thm:bisim}}.
\bpr
Let $u \in S'_Q$, $r \in S_P$, and  $u \bsim r$.
We must show that all three clauses of
Definition~\ref{def:bisim} hold, that is:
\bn

\item if $u \bsim r$ then $u \pj \AP = r \pj \AP$

\item if $u \bsim r$ and $(u,i,v) \in R_Q$ then
           $\ex s: (r,i,s) \in R_P \land v \bsim s$

\item if $u \bsim r$ and $(r,i,s) \in R_P$ then
           $\ex v: (u,i,v) \in R_Q \land v \bsim s$

\en

Clause~\ref{def:bisim:ap} holds by virtue of
clause~\ref{def:correctness-sim:ap} of
Definition~\ref{def:correctness-sim}.\\

\textit{Proof of clause~\ref{def:bisim:left-trans}}.
Assume $(u,i,v) \in R_Q$, and let $\inproc(u) = c$.
We show that there exists $s$ such that
$(r,i,s) \in R_P$ and $v \bsim s$.
By our construction of $P$ above, the transition $(u,i,v)$ generates
the arc $ARC_i^{u,v}$ in $P_i$. By
definition, the guard $B_i^{u,v}$ of $ARC_i^{u,v}$ is
\bleqn{(a)}
     $(\last = c \;\land\;
       \AND_{j \ne i} \stof{u \pj j} \;\land\;
       (\AND_{x \in \SH} x_{ci}^c = u(x)))$.
\eleqn
Now by Definition~\ref{def:correctness-sim} and $u \bsim r$, we have
$\inproc(u) = r(\last)$.
Hence $r \sat \last = c$.
Also by Definition~\ref{def:correctness-sim} and $u \bsim r$, we have
$u \pj \AP \; =\; r \pj \AP$.
Hence $r \sat \AND_{j \ne i} \stof{u \pj j}$.
Again by Definition~\ref{def:correctness-sim} and $u \bsim r$, we have
$\AND_{x \in \SH} r(\last) = c \imp u(x) = r(x_{ci}^{c}$.
Hence
$\AND_{x \in \SH,} u(x) = r(x_{ci}^{c})$. And so
$r \sat (\AND_{x \in \SH} x_{cj}^c = u(x))$.

Since $r$ satisfies all three conjuncts of (a), it
follows that the guard of 
$ARC_i^{u,v}$ is true in state $r$, and  therefore $ARC_i^{u,v}$
is enabled in $r$.
By Proposition~\ref{prop:invariants} and inspection of the action $A_i^{u,v}$
of $ARC_i^{u,v}$, executing of $ARC_i^{u,v}$ leads to a state $s$ such that
\bleqn{}
     $s(\last) = i$ and 
     $s \pj \AP = v \pj \AP$ and
     $(\AND_j x_{ij}^i = v(x))$.
\eleqn
By Definition~\ref{def:correctness-sim}, we have $v \bsim s$, as required.\\

\textit{Proof of clause~\ref{def:bisim:right-trans}}.
Assume $(r,i,s) \in R_P$. We show that there exists $v$ such that
$(u,i,v) \in R_Q$ and $v \bsim s$.

By our construction of $P$ above, the transition $(r,i,s)$ results
from executing an arc $ARC_i^{w,v}$ in $P_i$, for some $w,v$.
Let $\inproc(w) = c$.
By definition of $ARC_i^{w,v}$, we have  $r \sat \AND_{j \ne i} \stof{w \pj j}$,
and also $r \pj i = w \pj i$.
Hence, by the definition of $\stof{w}$ (Definition~\ref{def:stof}), 
$r \pj \AP = w \pj \AP$.
Also by definition of $ARC_i^{w,v}$, we have 
$r(\last) = \inproc(w) = c \land (\AND_{x \in \SH} r(x_{ci}^c) = w(x))$. 
Hence:
\bleqn{(b)}
   $r(\last) = \inproc(w) = c$ and
   $r \pj \AP = w \pj \AP$ and 
   $(\AND_{x \in \SH} r(x_{ci}^c) = w(x))$.
\eleqn
Since $u \bsim r$, we have
\bleqn{}
   $r(\last) = \inproc(u)$ and 
   $u \pj \AP = r \pj \AP$ and 
   $(\AND_{x \in \SH} r(x_{\last,i}^\last) = u(x))$.
\eleqn
From (b), $r(\last) = c$. Hence
\bleqn{(c)}
   $r(\last) = \inproc(u)$ and 
   $u \pj \AP = r \pj \AP$ and 
   $(\AND_{x \in \SH} r(x_{ci}^c) = u(x))$.
\eleqn
From (b,c) we have
\bleqn{(d)}
   $\inproc(w) = \inproc(u)$ and 
   $w \pj \AP = u \pj \AP$ and 
   $(\AND_{x \in \SH} w(x) = u(x))$.
\eleqn
Since all global states differ in either some atomic
proposition or some shared variable, or some incoming transition, by
Proposition~\ref{prop:unique-global}, 
we conclude from (d) that $w = u$.

By Proposition~\ref{prop:invariants} and inspection of the action
$A_i^{u,v}$ of $ARC_i^{u,v}$, executing  $ARC_i^{u,v}$
can only lead to a state $s$ such that
\bleqn{}
     $s(\last) = i$ and 
     $s \pj \AP = v \pj \AP$ and
     $(\AND_j x_{ij}^i = v(x))$.
\eleqn
By Definition~\ref{def:correctness-sim}, we have $v \bsim s$, as required.
\epr

\bco
\label{cor:bisim}
$M'_Q \sim M_P$.
\eco
\bpr
From Definition~\ref{def:correctness-sim} and our definition of the
initial states of $P$, we see that for every initial state $u_0$ of
$M'_Q$, there exists an initial state $r_0$ of $M_P$ such that 
$u_0 \bsim r_0$, and vice-versa. The result then follows from 
Theorem~\ref{thm:bisim} and Definition~\ref{def:bisim}.
\epr

\subsection{Phase Three}
\label{sec:phase-three}


We now express $ARC_i^{u,v}$ 
in a form that complies with
Definition~\ref{def:pairwise}, that is, as 
$\gc_{j \in I(i)} 
     \gd_{\l \in \{1,\ldots,n_j\}} \CB{i}{j}{\l} \ar \CA{i}{j}{\l}$,
where
$\CB{i}{j}{\l}$ can reference only variables in $\SH_{ij}$
    and atomic propositions in $\AP_j$,  and 
$\CA{i}{j}{\l}$ can update only variables in $\SH_{ij}$.
Recall that 
 $ARC_i^{u,v} = (u \pj i, B_i^{u,v} \ar A_i^{u,v}, v \pj i)$.
For the rest of this section, let $\inproc(u) = c$.
First consider $B_i^{u,v}$. By definition 
$B_i^{u,v} =
       (\last = c) \;\land\;
       \AND_{j \ne i} \stof{u \pj j} \;\land\;
       (\AND_{x \in \SH} x_{ci}^c = u(x))$.
Now $\stof{u \pj j}$ is a propositional formula over $\AP_j$, and so 
$\AND_{j \ne i} \stof{u \pj j}$
is a conjunction of propositional formulae over $\AP_j$, and so it poses no problem.
Likewise, since $(\AND_{x \in \SH} x_{ci}^c = u(x))$
is a conjunction over pairwise shared variables, it also is
unproblematic.
$\last = c$ is not in the pairwise form given above since it refers
to the ghost variable $\last$. Note that $\inproc(u)$ is a constant,
and so is not problematic in this regard.

Now $\last = c$ checks that the last
process to execute is $P_c$.  In terms of timestamps, it is equivalent
to $\AND_{j \ne c} t_c^j \tgt t_j^c$, i.e., 
$P_c$ has executed more recently than all other processes.
However, the timstamps $t_j^c$ are inaccessible to $P_i$, 
and the $t_c^j$ are accessible to $P_i$ only in the special case that
$c=i$, which does not hold generally.
The purpose of the timestamp vectors is precisely to deal with this 
problem. Recall that 
$tv_{ci}^c.j$ is maintained equal to $t_c^j$, and 
$tv_{ji}^j.c$ is maintained equal to $t_j^c$.
Hence, we replace  $\last = c$ by the equivalent
\bleqn{(*)}
            $\AND_{j \ne c} tv_{ci}^c.j \tgt  tv_{ji}^j.c$. 
\eleqn
which moreover can be evaluated by $P_i$,
since it refers only to timestamp vectors that are accessible to $P_i$.

Now the expression $tv_{ci}^c.j \tgt  tv_{ji}^j.c$ refers to 
$tv_{ci}^c$, which is shared by $P_c$ and $P_i$, and 
$tv_{ji}^j$, which is shared by $P_j$ and $P_i$.
Thus it is not in pairwise form.
We fix this as follows.
$tv_{ci}^c.j \tgt tv_{ji}^j.c$ is equivalent to 
$(tv_{ci}^c.j = 0 \land tv_{ji}^j.c = 1) \lor
 (tv_{ci}^c.j = 1 \land tv_{ji}^j.c = 2) \lor
 (tv_{ci}^c.j = 2 \land tv_{ji}^j.c = 0)$, by definition of $\tgt$.
Hence, $(*)$ is equivalent to\\[2ex]
$\AND_{j \ne c} (tv_{ci}^c.j = 0 \land tv_{ji}^j.c = 1) \lor
                (tv_{ci}^c.j = 1 \land tv_{ji}^j.c = 2) \lor
                (tv_{ci}^c.j = 2 \land tv_{ji}^j.c = 0)$.\\[2ex]
This formula has length in $O(K)$.
We convert this to disjunctive normal form, resulting in a 
formula of length in $O(exp(K))$.
Let the result be $D_1 \lor \ldots \lor D_n$ for some $n$.
Each $D_m$, $1 \le m \le n$ is a conjunction of literals, where each
literal has one of the forms $(tv_{ci}^c.j ~op~ ts)$, $(tv_{ji}^j.c ~op~ ts)$,
where $op \in \{=, \ne\}$, and $ts \in \{0,1,2\}$.
Specifically,
\bleqn{}
$D_m = LIT_m^c(tv_{ci}^c.j) \land \AND_{j \not\in \set{c,i}} LIT_m^j(tv_{ji}^j.c)$,
\eleqn
where 
$LIT_m^c(tv_{ci}^c.j)$ is a conjunction of literals of the form $tv_{ci}^c.j ~op~ ts$, and 
$LIT_m^c(tv_{ji}^j.c)$ is a conjunction of literals of the form $tv_{ji}^j.c ~op~ ts$.
Moreover, since logical equivalence to (*) has been maintained, we have
\bleqn{}
       $(D_1 \lor \ldots \lor D_n) \equiv (\last = c)$.
\eleqn
For $m \in \set{1,\ldots,n}$, define:
\bleqn{}
       $B_i^{u,v}(m) \;\df\;
        D_m \land  \AND_{j \ne i} \stof{u \pj j} \;\land\;
       (\AND_{x \in \SH} x_{ci}^c = u(x))$
\eleqn
where we abuse notation by using $B_i^{u,v}$ as the name for the ``array''
of guards $B_i^{u,v}(m)$, and also as the name for the guard of 
$ARC_i^{u,v}$, as defined above. The use of the index $(m)$ will
always disambiguate these two uses.  

We now define the set of arcs $ARCS_i^{u,v}$ to contain $n$ arcs,
$a(1),\ldots,a(n)$, where
\bleqn{}
    $a(m) \;\df\; 
       (u \pj i,
         B_i^{u,v}(m) \ar A_i^{u,v},
        v \pj i)$
\eleqn
for all $m \in 1,\ldots,n$.
In particular, all these arcs
start in local state $u \pj i$ of $P_i$ and end in local
state $v \pj i$ of $P_i$.

\bp
\label{prop:guard}
$(\OR_{1 \le m \le n} B_i^{u,v}(m)) \equiv B_i^{u,v}$
\ep
\bpr Immediate from the definitions and distribution of $\land$
through $\lor$.
\epr

It remains to show how each $a(m)$ can be rewritten into pairwise
normal form.
For all $j \not\in \set{i,c}$,
define 
\bleqn{}
      $B_i^{u,v}(m,j)  \df LIT_m^j(tv_{ji}^j.c) \land \stof{u \pj j}$
\eleqn
For $j = c$.
\bleqn{}
      $B_i^{u,v}(m,c) \df
		LIT_m^c(tv_{ci}^c.j) \land 
		\stof{u \pj c} \;\land\;
	        (\AND_{x \in \SH} x_{ci}^c = u(x))$
\eleqn
Note that this works for both $c \ne i$ and $c = i$. The case $c=i$ is
why we needed to allow  $x_{ii}^i$ and $tv_{ii}^i$. Otherwise we would
need a special case to deal with $c=i$.
In effect, when $c=i$ we include $B_i^{u,v}(m,c)$ as a conjunct of
$B_i^{u,v}(m,\l)$, where $P_\l$ is the process arbitrarily chosen to ``share'' 
$x_{ii}^i$ and $tv_{ii}^i$ with $P_i$.
This allows us to conform to pairwise normal form, and use 
$(\AND_{j \ne i} B_i^{u,v}(m,j))$ as the guard of the arc:

\bp
\label{prop:guardc}
$(\AND_{j \ne i} B_i^{u,v}(m,j)) \equiv B_i^{u,v}(m)$
\ep

\textit{Proof of Proposition~\ref{prop:guardc}}.
\bpr
by definition, 
       $B_i^{u,v}(m) =
        D_m \land  \AND_{j \ne i} \stof{u \pj j} \;\land\;
       (\AND_{x \in \SH} x_{ci}^c = u(x))$.
We also have, by construction,
$D_m = LIT_m^c(tv_{ci}^c.j) \land \AND_{j \not\in \set{c,i}} LIT_m^j(tv_{ji}^j.c)$.
Hence
$B_i^{u,v}(m)  \equiv
	LIT_m^c(tv_{ci}^c.j) \land 
        (\AND_{j \not\in \set{c,i}} LIT_m^j(tv_{ji}^j.c)) \land 
	(\AND_{j \ne i} \stof{u \pj j}) \land
       (\AND_{x \in \SH} x_{ci}^c = u(x))$.

Splitting up conjunctions and rearranging gives us:

$B_i^{u,v}(m)  \equiv
	(\AND_{j \not\in \set{c,i}} LIT_m^j(tv_{ji}^j.c)) \land 
	(\AND_{j \not\in \set{c,i}} \stof{u \pj j}) \land
	LIT_m^c(tv_{ci}^c.j) \land 
	\stof{u \pj c} \land
	(\AND_{x \in \SH} x_{c,i}^c = u(x))$.

Grouping together the first two conjunctions, and the last three:

$B_i^{u,v}(m)  \equiv
	(\AND_{j \not\in \set{c,i}} LIT_m^j(tv_{ji}^j.c) \land \stof{u \pj j}) \land 
	[LIT_m^c(tv_{ci}^c.j) \land 
	\stof{u \pj c} \land
	(\AND_{x \in \SH} x_{c,i}^c = u(x))]$.

Now $LIT_m^j(tv_{ji}^j.c) \land \stof{u \pj j}$ is just $B_i^{u,v}(m,j)$, and 
$[LIT_m^c(tv_{ci}^c.j) \land 
	\stof{u \pj c} \land
	(\AND_{x \in \SH} x_{c,i}^c = u(x))]$
is just $B_i^{u,v}(m,c)$.
Hence

$B_i^{u,v}(m)  \equiv
	(\AND_{j \not\in \set{c,i}} B_i^{u,v}(m,j)) \land B_i^{u,v}(m,c)$.
Thus
$B_i^{u,v}(m)  \equiv
	\AND_{j \ne i} B_i^{u,v}(m,j)$.
\epr


The timestamps $t_i^j$ are written and read by $P_i$ and no other process.
To achieve pariwise normal form, we now make the $t_i^j$ part of the local state
of $P_i$. Thus, we replace each local state $r_i$ of $P_i$ by $3^K$ local
states, each of which agrees with $r_i$ on the atomic propositions in $\AP_i$.
There is one such state for every different assignment of timestamp values to
$t_1^1,\ldots,t_1^K$.
Call the new process that results $PP_i$, and let 
$PP = (St, PP_1 \pl \cdots \pl PP_K)$.
Note that $PP$ has the same initial states as $P$.
Let $r'_i$ be a local state of $PP_i$,
and let $t_i^1,\ldots,t_i^K$ have some values $d_1,\ldots,d_K$ in $r'_i$.
Likewise let $s'_i$ agree with $s_i$ on the atomic propositions in $\AP_i$, and
let  $t_1^1,\ldots,t_1^K$ have some values $d'_1,\ldots,d'_K$ in $s'_i$.
Then, the set of arcs $ARCS_i^{u,v}(r'_i,s'_i)$ is defined as follows.

$ARCS_i^{u,v}(r'_i,s'_i)$ contain $n$ arcs, $a'(1),\ldots,a'(n)$,  where
    $a'(m) \;\df\;$\linebreak
       $(r'_i,
        \gc_{j \ne i} BB_i^{u,v}(m,j) \ar AA_i^{u,v}(m,j),
        s'_i)$
for all $m \in 1,\ldots,n$.
In particular, all these arcs start in $r'_i$ and end in $s'_i$.
Also:\\
For all $j \ne i$,
\bleqn{}
   $BB_i^{u,v}(m,j) \df
	B_i^{u,v}(m)_i^j \land step(d_j,tv_{ji}^j.i) = d'_j$
\eleqn
For all $j \ne i$,
\bleqn{}
  $AA_i^{u,v}(m,j) \;\df\;
         (tv_{ij}^i := \tpl{\ldots,step(d_j,tv_{ji}^j.i),\ldots};\ 
          \pl_{x \in \SH} \ x_{ij}^i := v(x))$
\eleqn
The new conjunct $step(d_j,tv_{ji}^j.i) = d'_j$ in effect checks that the values
of the timestamps $t_i^j$ for all $j$ in the new local states are exactly
those that the operation $step(t_i^j,t_j^i)$ would return, i.e., those values
that would indicate that $P_i$ has excecuted later than $P_j$.
The timestamp vector $tv_{ij}^i$ can now be updated correctly without violating
pairwise normal form, since the update can be performed using the $d_j$ values,
which are constants, and the $tv_{ji}^j.i$. which are shared pairwise between
$P_i$ and $P_j$, and are therefore permitted by pairwise normal form.

Let $M_{PP} = (St_P, S_P, R_{PP})$ be the state-transition diagram of $PP$.
Note that $PP$ and $P$ have the same initial states, and the same global states, by
definition. 

\bt
\label{thm:equal-MP-MPP}
$M_{P} \sim M_{PP}$
\et

\textit{Proof of Theorem~\ref{thm:equal-MP-MPP}}
\bpr
Let $(r,i,s) \in R_P$.  
$(r,i,s)$ results from executing an arc $ARC_i^{u,v}$.
Hence $B_i^{u,v}$ is true in state $r$.
By Proposition~\ref{prop:guard}, some $B_i^{u,v}(m)$ is true in state $r$.
Hence $\AND_{j \ne i} B_i^{u,v}(m,j)$ is true in state $r$, by 
Proposition~\ref{prop:guardc}.

Now let $r',s'$ be the states in $M_{PP}$ that correspond to states $r,s$ in
$M_P$, that is $r'$ and $r$ agree on all atomic propositions and shared
variabled (including timestamps) and likewise $s$ and $s'$.

Let $r'_i = r' \pj i$, $s'_i = s' \pj i$.
Let $t_i^1,\ldots,t_i^K$ have values $d_1,\ldots,d_K$ in $r'_i$ (and hence also
in $r'$),
and values $d'_1,\ldots,d'_K$ in $s'_i$ (and hence also in $s'$).
($r,r'$ are essentially different ways of refereeing to the same state, 
to indicate whether the containing structure is $M_P$ or $M_{PP}$, and likewise
$s,s'$).

Since $(r,i,s)$ results from executing $ARC_i^{u,v}$, 
$step(d_j,tv_{ji}^j.i) = d'_j$ must hold, since 
the action $A_i^{u,v}$ of $ARC_i^{u,v}$ contains the assignment 
$\pl_{j \ne i} \ t_i^j := \step(t_i^j, tv_{ji}^i.j)$.
Hence $\AND_{j \ne i} BB_i^{u,v}(m,j)$ is true in state $r'$.
Thus, arc $a'(m)$ of the set $ARCS_i^{u,v}(r'_i,s'_i)$ is enabled in state $r'$.
Execution of  $a'(m)$ in state $r'$ leads to state $s'$, by definition of 
$AA_i^{u,v}(m,j)$. 
Hence $(r',i's') \in R_{PP}$.

Now let $(r',i,s') \in R_{PP}$.  
$(r',i,s')$ results from executing an arc $a'(m)$ of some set
$ARCS_i^{u,v}(r'_i,s'_i)$,
where $r'_i = r' \pj i$, $s'_i = s' \pj i$.
We can run the previous argument ``backwards'' to show that 
$ARC_i^{u,v}$ is enabled in state $r$ of $M_P$, and its execution results in
state $s$ of $M_P$. Hence $(r,i,s) \in R_P$.

We have in fact showed that $R_P = R_{PP}$, i.e., that the structures $M_P$ and
$M_{PP}$ are identical. Hence they are certainly bisimilar.
\epr

\bco
\label{cor:bisim-MQ-MPP}
$M_Q \sim M_{PP}$
\eco
\bpr
Immediate from Proposition~\ref{prop:unique-correct},
Corollary~\ref{cor:bisim} and Theorem~\ref{thm:equal-MP-MPP}, along with the
transitivity of bisimulation.
\epr

Since $PP$ is in pairwise normal form by construction, our
main result follows immediately:
\bt
\label{thm:main}
Let $Q$ be any finite-state concurrent program.
Then there exists a concurrent program $PP$ such that
(1) the global state transition diagrams of $Q$ and $PP$ are bisimilar, and
(2) $PP$ is in pairwise normal form.
\et

Our result shows that $PP$ and $Q$ have essentially the same behavior,
since strong bisimulation is the strongest notion of equivalence
between concurrent programs.
A consequence of our result is that $PP$ and $Q$ satisfy the same
specifications, for many logics of programs.
Recall that $M_{PP}$ and $M_Q$ are the global state
transition diagrams of $P$ and $Q$, respectively.
Let $f$ be a formula of the temporal logic $\CTLS$
\cite{Em90}, and define 
$M_Q, u \sat f$ to mean $\fa u \in St_Q: M_Q, u \sat f$, and
$M_{PP}, s \sat f$ to mean $\fa s \in St_P: M_P, s \sat f$, 
where
 $M_Q, u \sat f$ and $M_{PP}, s \sat f$ refer to the 
usual satisfaction relation of $\CTLS$ \cite{Em90}.
Then we have:

\bco
Let $f$ be a formula of $\CTLS$. Then $M_Q \sat f$ iff $M_{PP} \sat f$.
\eco
\bpr
Immediate from Corollary~\ref{cor:bisim-MQ-MPP}
and Theorem 14 in \cite[chapter 11]{CGP99}.
\epr

We could easily establish similar results for other logics, such as
the mu-calculus.

\subsection{Complexity Results}
\label{sec:complexity}

For a single process $Q_i$, define $|Q_i|$, the size of $Q_i$, to be the
size of the representation of $Q_i$ using a standard
complexity-theoretic encoding, i.e., enumeration for sets, 
character strings for guards and actions etc.
Likewise define $|PP_i|$.
Define $|Q|$, the size of $Q$, to be $|St_Q|$ + $|Q_1| + \cdots + |Q_K|$,
and $|PP|$, the size of $PP$, to be $|St_P|$ + $|PP_1| + \cdots + |PP_K|$.

Define the size of a Kripke structure to be the number of states plus
the number of transitions.

\bt
\label{thm:size}
$|PP|$ is in $O(K exp(|Q|+K))$. 
\et
\bpr
$|M_Q|$ is in $O(exp(|Q|))$ by Definition~\ref{def:gstd}.
$|M'_Q|$ is in $O(K \cdot |M_Q|)$, since each state and transition in
$M_Q$ is ``replicated'' at most $K$ times.
So $|M'_Q|$ is in $O(K exp(|Q|))$.

For each transition in $M'_Q$, 
$PP$ contains a number of arcs that is in $O(exp(K))$.
Hence $|PP|$ is in $O(|M'_Q| \cdot exp(K))$, and so
$|PP|$ is in $O(K \cdot exp(|Q|) \cdot exp(K))$. Thus
$|PP|$ is in $O(K exp(|Q|+K))$.
\epr


\section{Related Work}
\label{sec:related}

It has been long known that a multiple-reader multiple writer atomic
register can be implemented using a set of single-reader single-writer
registers, and three are many such atomic register constructions in
the literature \cite[chapter 10]{AW98}.  Since, by definition, a
single-reader single-writer register is shared by two processes, these
constructions may seem to subsume our result.  However, the atomic
register constructions do not respect pairwise normal form. For
example, they may involve the operation of taking the maximum
over a set of single-reader single-writer 
registers that involve many different pairs of processes.
This direct use of register values corresponding to many different
pairs, in computing a single expression value, is a direct violation
of pairwise normal form. 


%
\section{Conclusions and Future Work}
\label{sec:conc}

We showed that any finite-state shared memory concurrent program can
be rewritten in pairwise normal form, up to strong bisimulation, for a
high-atomicity model of concurrent computation.  A topic of future
work is to establish a similar result in a low-atomicity model, for
example that presented in \cite{AE01}.  Our results have significant
implications for the efficient synthesis and model-checking of
finite-state shared memory concurrent programs. In particular, they
show that the approaches of \cite{Att99a,AE98,AC04a} do not sacrifice
any expressive power by restricting attention to pairwise normal form.


\small

\bibliographystyle{plain}
\bibliography{BIBFILES/ABBREV,BIBFILES/DIST,BIBFILES/IOAUT,BIBFILES/MODEL,BIBFILES/SYNTH,BIBFILES/LOGIC,BIBFILES/OS}

\end{document}

%% file: include/macros-tr.tex
\newcommand{\be}{\begin{itemize}}
\newcommand{\ee}{\end{itemize}}
\newcommand{\bn}{\begin{enumerate}}
\newcommand{\en}{\end{enumerate}}

\newcommand{\bp}{\begin{proposition}}
\newcommand{\ep}{\end{proposition}}
\newcommand{\bl}{\begin{lemma}}
\newcommand{\el}{\end{lemma}}
\newcommand{\bco}{\begin{corollary}}
\newcommand{\eco}{\end{corollary}}
\newcommand{\bt}{\begin{theorem}}
\newcommand{\et}{\end{theorem}}
\newcommand{\bpr}{\begin{proof}}
\newcommand{\epr}{\end{proof}}
\newcommand{\bd}{\begin{definition}}
\newcommand{\ed}{\end{definition}}

\newcommand{\beqn}{\begin{centeqn}}
\newcommand{\eeqn}{\end{centeqn}}

\newcommand{\bleqn}[1]{\begin{centlabeqn}{#1}}
\newcommand{\eleqn}{\end{centlabeqn}}

\newcommand{\bc}{\begin{center}}
\newcommand{\ec}{\end{center}}

\newcommand{\bs}{\bigskip}
\renewcommand{\ss}{\smallskip}

\newcommand{\bfg}{\begin{figure}[t]}
\newcommand{\efg}{\end{figure}}

\newenvironment{centeqn}	
   {{\ss\\ \hspace*{\fill}}} 
   {\hspace*{\fill}\ss\\}

\newsavebox{\EqnLabel}
\newenvironment{centlabeqn}[1]		
   {\sbox{\EqnLabel}{#1}
    {\bs\\ \hspace*{\fill}}
   } 
   {\hfill{\makebox[0in][r]{\usebox{\EqnLabel}}}\bs\\}

\newenvironment{centeqn-nbsp} 
   {{\ms\\ \hspace*{\fill}}}
   {\hspace*{\fill}}

\newcommand{\intrdef}[1]{\emph{#1\/}}   
\newcommand{\intr}[1]{\emph{#1\/}}      

\newcommand{\empb}[1]{\textbf{#1\/}}

\newlength{\parboxwidth}
\newcommand{\ba}{\bigwedge}	
\newcommand{\AND}{\bigwedge}

\newcommand{\bigor}{\bigvee}	
\newcommand{\OR}{\bigvee}
\newcommand{\UN}{\bigcup}

\newcommand{\ar}{\rightarrow}	

\newcommand{\df}{\mbox{$\:\stackrel{\rm df}{=\!\!=}\:$}}

\newcommand{\ex}{\exists}
\newcommand{\fa}{\forall}

\newcommand{\imp}{\Rightarrow}		

\newcommand{\ints}{\cap}
\renewcommand{\l}{\ell}

\newcommand{\oneton}{[n]}
\newcommand{\onetok}{[K]}

\newcommand{\pl}{\!\parallel\!}

\newcommand{\sat}{\models}

\newcommand{\sub}{\subseteq}
\newcommand{\un}{\cup}
\newcommand{\up}{\pj}

\newcommand{\set}[1]{\{ #1 \}}
\newcommand{\setinsert}[2]{#1 \gets #1 \cup \{#2\}}
\newcommand{\setdelete}[2]{#1 \gets #1 -  \{#2\}}
\newcommand{\tpl}[1]{\langle #1 \rangle}
	
\newcommand{\pj}{\raisebox{0.2ex}{$\upharpoonright$}}

\newcommand{\gc}{\otimes}	
\newcommand{\gd}{\oplus}	

\newcounter{levelone}

\newcounter{leveltwo}

\newenvironment{proof}{\textit{Proof.} }	
                      {\hfill{$\Box$}}

\newcommand{\trans}{\mathrm{TRANSFORM}}
\newcommand{\tls}{<_o}
\newcommand{\tgt}{>_o}
\newcommand{\last}{\mathit{last}}
\newcommand{\step}{\mathit{step}}
\newcommand{\bsim}{\mathit{\,\bowtie\,}}
\newcommand{\inprocs}{\mathit{in\_procs}}
\newcommand{\inproc}{\mathit{in}}

\newcommand{\IF}{\mbox{$\mathbf{if}$}\xspace}
\newcommand{\THEN}{\mbox{$\mathbf{then}$}\xspace}
\newcommand{\ELSE}{\mbox{$\mathbf{else}$}\xspace}
\newcommand{\ENDIF}{\mbox{$\mathbf{endif}$}\xspace}

\newcommand{\FORALL}{\mbox{$\mathbf{for all}$}\xspace}
\newcommand{\DO}{\mbox{$\mathbf{do}$}\xspace}

\newcommand{\ENDFOR}{\mbox{$\mathbf{endfor}$}\xspace}

\newcommand{\REPEAT}{\mbox{$\mathbf{repeat}$}\xspace}

\newcommand{\ENDREPEAT}{\mbox{$\mathbf{endrepeat}$}\xspace}

\newcommand{\RETURN}{\mbox{$\mathbf{return}$}\xspace}

\newcommand{\stof}[1]{\{\hspace*{-0.3em}|#1|\hspace*{-0.3em}\}}

\newcommand{\CA}[3]{\mbox{$A_{#1,#3}^{#2}$}}  
\newcommand{\CB}[3]{\mbox{$B_{#1,#3}^{#2}$}}  

\newcommand{\PP}[2]{\mbox{$P_{#1}^{#2}$}}   

\newcommand{\mv}[2]{\mbox{$a_{#1}^{#2}$}}  

\newcommand{\AP}{\mbox{$\cal AP$}}  
\newcommand{\CTLS}{\mbox{$\mathrm{CTL}^*$}}   	

\newcommand{\SH}{\mathcal{SH}}

\newcommand{\lfalse}{\mathit{false}}
\newcommand{\ltrue}{\mathit{true}}

\newtheorem{theorem}{Theorem}
\newtheorem{lemma}[theorem]{Lemma}
\newtheorem{proposition}[theorem]{Proposition}
\newtheorem{corollary}[theorem]{Corollary}

\newtheorem{definition}{Definition}

\newcommand{\ms}[1]{%
        \relax\ifmmode
                \mathord{\mathcode`\-="702D\it #1\mathcode`\-="2200}%
        \else
                {\it #1}%
        \fi
}

\newcommand{\horline}{\rule{\textwidth}{1pt}}

%% file: express-tr.bbl
\newcommand{\SortNoop}[1]{}
\begin{thebibliography}{10}

\bibitem{Att99a}
P.~C. Attie.
\newblock Synthesis of large concurrent programs via pairwise composition.
\newblock In {\em {CONCUR}'99: 10th International Conference on Concurrency
  Theory}, number 1664 in LNCS. Springer-Verlag, Aug. 1999.

\bibitem{AE98}
P.~C. Attie and E.~A. Emerson.
\newblock Synthesis of concurrent systems with many similar processes.
\newblock {\em ACM Trans. Program. Lang. Syst.}, 20(1):51--115, Jan. 1998.

\bibitem{AE01}
P.~C. Attie and E.~A. Emerson.
\newblock Synthesis of concurrent systems for an atomic read/write model of
  computation.
\newblock {\em ACM Trans. Program. Lang. Syst.}, 23(2):187--242, Mar. 2001.
\newblock Extended abstract appears in Proceedings of the 15'th ACM Symposium
  of Principles of Distributed Computing ({PODC}), Philadelphia, May 1996,
  111--120.

\bibitem{Att03a}
P.C. Attie.
\newblock Synthesis of large dynamic concurrent programs from dynamic
  specifications.
\newblock Technical report, Northeastern University, Boston, MA, 2003.
\newblock {A}vailable at \url{http://www.ccs.neu.edu/home/attie/pubs.html}.

\bibitem{AC04a}
P.C. Attie and H.~Chockler.
\newblock Efficiently verifiable sufficient conditions for deadlock-freedom of
  large concurrent programs.
\newblock Technical report, Northeastern University, Boston, MA, 2004.
\newblock {A}vailable at \url{http://www.ccs.neu.edu/home/attie/pubs.html}.

\bibitem{AW98}
H.~Attiya and J.~Welch.
\newblock {\em Distributed Computing}.
\newblock McGraw Hill, London, UK, 1998.

\bibitem{CGP99}
E.M. Clarke, O.~Grumberg, and D.A. Peled.
\newblock {\em Model Checking}.
\newblock MIT Press, Cambridge, MA, 1999.

\bibitem{DS97}
Dolev D. and Shavit N.
\newblock Bounded concurrent time-stamping.
\newblock {\em SIAM J. Comput.}, 26(2):418--455, Apr. 1997.

\bibitem{Dij76}
E.~W. Dijkstra.
\newblock {\em A Discipline of Programming}.
\newblock Prentice-Hall Inc., Englewood Cliffs, N.J., 1976.

\bibitem{Em90}
E.~A. Emerson.
\newblock Temporal and modal logic.
\newblock In J.~Van Leeuwen, editor, {\em Handbook of Theoretical Computer
  Science}, volume {B}, \textit{Formal Models and Semantics}. The MIT
  Press/Elsevier, Cambridge, Mass., 1990.

\bibitem{EC82}
E.~A. {\SortNoop{EmersonB, E. A.}}~Emerson and E.~M. Clarke.
\newblock Using branching time temporal logic to synthesize synchronization
  skeletons.
\newblock {\em Sci. Comput. Program.}, 2:241 -- 266, 1982.

\end{thebibliography}
